\documentclass[slac_one]{revtex4}
\usepackage{graphicx}
\usepackage{fancyhdr}
\begin{document}

\title{{\small{Hadron Collider Physics Symposium (HCP2008),
Galena, Illinois, USA}}\\ 
\vspace{12pt}
Measurements of B Hadron Properties at the Tevatron}

%

\author{M. Hartz (on behalf of the D0 and CDF Collaborations)}
\affiliation{University of Pittsburgh, Pittsburgh, PA 15260, USA}
%

\begin{abstract}
Recent measurements of $B$ hadron properties carried out in $\sqrt{s}=1.96$ TeV $p\bar{p}$ collisions at the 
Tevatron are reviewed.  Included are measurements of the $B_c^{+}$ meson lifetime, using $J/\psi+l+X$ final states, and
the mass, using $J/\psi+\pi$ final states, a flavor specific measurement of the $B_s^0$ lifetime in $D_s+\pi+X$ decays, 
simultaneous measurements of $\tau_{B_s}$ and $\Delta\Gamma_{B_s}$ in $J/\psi+\phi$ decays, the first direct 
evidence and mass measurements of the $\Xi_b^-$ baryon, and measurements of the polarization of 1S charmonium.
For all measurements, charge conjugate modes are included.
\end{abstract}

\maketitle

\thispagestyle{fancy}


\section{INTRODUCTION} 
The measured properties of hadrons containing heavy quarks, $b$ and $c$, test
the predictions of quantum chromodynamics (QCD) in bound states that probe 
kinematic regions not seen in light hadrons.  To carry out calculations, effective
field theories have been developed for systems including a single heavy quark,
heavy quark effective theory (HQET)~\cite{Eichten:1989zv}, and two heavy quarks, nonrelativistic 
QCD (NRQCD)~\cite{Quigg:1979vr}.  A number of tools have been adapted to work
in the context of these effective field theories including lattice QCD (lQCD)~\cite{Thacker:1990bm}, 
QCD sum rules~\cite{Shifman:1978bx}, and the operator product expansion (OPE)~\cite{Brodsky:1980ny}.  
The measured properties presented in this paper help in constraining some of the approaches
 used within the effective field theories.

The Tevatron, a $\sqrt{s}=1.96$ TeV $p\bar{p}$ collider, provides collisions for
two multipurpose detectors, the D0 detector and the upgraded Collider Detector at Fermilab (CDF II).  
Pairs $b\bar{b}$ production is the dominant source of $b$ quarks with a cross section of $\sim30$ $\mu$b~\cite{Acosta:2004yw}
compared to a total inelastic cross section of $\sim 50$ mb. All measurements presented in this 
paper use between $0.8$ and $2.8$ fb$^{-1}$ of integrated luminosity of collisions measured by the 
D0 and CDF II detectors.

Throughout this paper charge conjugate modes are implied unless noted otherwise.

\section{PROPERTIES OF THE $B_c^{+}$ MESON}
The $B_c^{+}$ meson, the ground state meson consisting of a $b$ and $c$ quark, is unique
among mesons since it both contains two heavy mesons and decays weakly.  The doubly heavy 
nature of the $B_c^{+}$ allows for the application of NRQCD while calculating its properties,
including the singular application of NRQCD to weak decay properties. 

The $B_c^{+}$ decays through three tree level processes: decay of the $c$ quark
through a $W$ leaving a $B$ meson in the final state, decay of the $b$ quark through a 
$W$ leaving charmonium in the final state, and annihilation of the $b$ and $c$ quarks 
leaving $\tau+\bar{\nu}_{\tau}$ or $c+\bar{s}$ in the final state.  While the decays of the $c$ 
quark should be expected to have the largest branching fractions, channels involving
the decay of the $b$ quark can be more useful from an experimental perspective since
the reconstruction of the $J/\psi$ through the dimuon mode takes place at a higher 
efficiency than any modes involving a $B_s^0$.  Table~\ref{bc_frac} compares 
predictions for branching fractions of the $B_c^{+}$ to final states with $J/\psi$ and $B_s^0$.

\begin{table}[t]
\begin{center}
\caption{Predicted $B_c^{+}$ branching fractions for decay channels that are candidates 
for experimental analysis.}
\begin{tabular}{|l|c|c|c|}
\hline
 & \multicolumn{3}{|c|}{ $B_c$ Branching Fraction($\%$)} \\ \hline
Decay Mode &~\cite{Kiselev:2002vz}&~\cite{Ivanov:2006ni}&~\cite{AbdElHady:1999xh} \\ \hline
$B_{s}^{0}+\pi^{\pm}$                & 16.4 & 3.9 & 1.56  \\ \hline
$B_{s}^{0}+e/\mu^{\pm}+\nu_{e/\mu}$  & 4.03 & 1.10 & 0.98  \\ \hline
$J/\psi+\pi^{\pm}$                   & 0.13 & 0.17 & 0.11  \\ \hline
$J/\psi+e/\mu^{\pm}+\nu_{e/\mu}$     & 1.9 & 2.07 & 1.44  \\ \hline
\hline
\end{tabular}
\label{bc_frac}
\end{center}
\end{table}

\subsection{Lifetime in $B_{c}^{+}\rightarrow J/\psi+l^{+}+X$ Decays}
NRQCD predicts a $B_c^{+}$ lifetime that is short compared to the $B^0/B^{+}$ lifetimes.  
Calculations using QCD sum rules predict a 
lifetime of $0.47\pm0.05$ ps~\cite{Kiselev:2000pp}, while calculations using the optical 
theorem and OPE predict $0.52$ ps~\cite{Beneke:1996xe} with a range of $0.4-0.7$ ps depending 
on the value of the charm quark mass.  The $B_c$ lifetime was first measured in CDF Run I data and
found to be $\tau_{B_c} = 0.46^{+0.18}_{-0.16}(stat.)\pm0.03(syst.)$ ps~\cite{Abe:1998wi}.

D0 and CDF measure the $B_c$ lifetime in the inclusive 
$B_{c}^{\pm}\rightarrow J/\psi(\mu\mu)+l^{+}+X$ decay mode, which is expected to be dominated by 
semileptonic decays to $J/\psi+l^{+}+\nu_{l}$ directly.
D0 uses 1.3 fb$^{-1}$ of $J/\psi+\mu+X$ events while CDF uses 1.0 fb$^{-1}$ of $J/\psi+l^++X$ where 
$l^+$ can be a muon or electron.  For both experiments, events are triggered on by the presences of the 
dimuon pair from the $J/\psi$ decay.

The unmeasured particles in the inclusive decays require a correction to account for the missing momentum
which can be modeled with a sample of simulated signal events and is defined as
\begin{equation}
K = \frac{p_{T}(J/\psi l)}{p(B_c)\cdot \hat{p}_T(J/\psi l)}
\end{equation} 
The missing momentum also implies a broad mass peak, so backgrounds are not estimated using mass sidebands.
Instead, background are modeled using data and and simulation and include: fake $J/\psi$ plus third lepton, true $J/\psi$
where the third lepton is faked by a hadron, true $J/\psi$ and leptons that do not originate from the 
same decay, and prompt $J/\psi$ plus a third lepton.   
The D0 measurement includes models of the mass distributions in a simultaneous fit of the lifetime
and mass distributions, while the CDF measurement fits the lifetime distribution only.

\begin{figure*}[t]
\centering
\includegraphics[width=100mm]{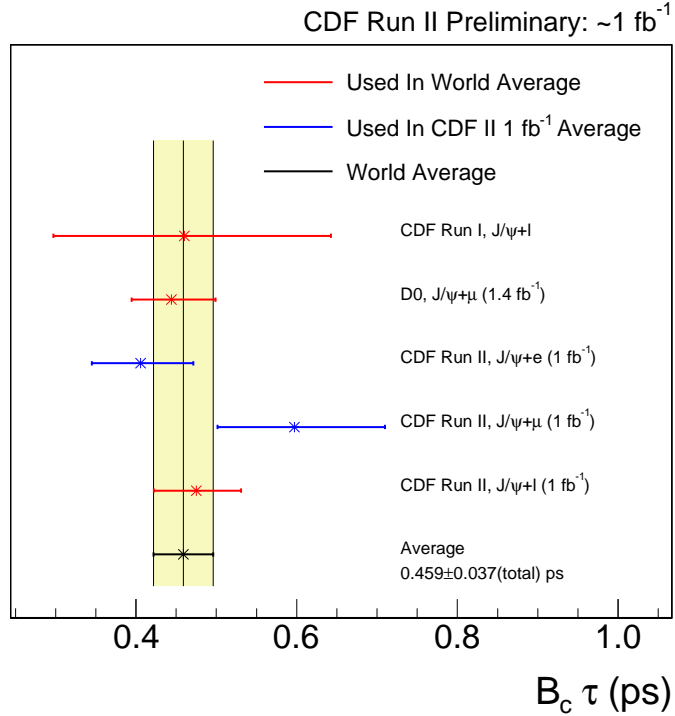}
\caption{Average of $B_c^{+}$ lifetime measurements from the Tevatron.  The world average is a weighted
average of results assuming no correlation in the uncertainties.} 
\label{bclt_sum}
\end{figure*}

D0 measures the lifetime with world best precision~\cite{:2008rb}: 
\begin{eqnarray*}
\tau_{B_c} = 0.448^{+0.038}_{-0.036}(stat.)\pm0.032(syst.) \: \mbox{ps}
\end{eqnarray*}
Systematic uncertainties limit further improvements in precision and are due to uncertainties in the 
mass and lifetime models used for the backgrounds. CDF measures the lifetime with similar precision~\cite{cdf_bclt}: 
\begin{eqnarray*}
\tau_{B_c} = 0.476^{+0.053}_{-0.049}(stat.)\pm0.018(syst.) \: \mbox{ps}
\end{eqnarray*}
For the CDF measurement, systematic uncertainties are limited by various tests of the background
models using data. The results are summarized in Fig.~\ref{bclt_sum} along with the CDF Run I measurement 
and a weighted average of 
\begin{eqnarray*}
\tau_{B_c} = 0.459\pm0.037 \: \mbox{ps}
\end{eqnarray*}
The average result is in good agreement with predictions from NRQCD and provides already constraining
information for theoretical predictions.

\subsection{Mass in $B_{c}^{+}\rightarrow J/\psi+\pi^{+}$ Decays}
The mass of the $B_{c}^{+}$ meson can be estimated in NRQCD using lQCD, 
$M=6304\pm12^{+18}_{-0}$ MeV/c$^2$~\cite{Brambilla:2001qk}, and potential models,
$M=6247-6286$ MeV/c$^2$~\cite{Godfrey:2004ya}.  The mass of the $B_c^{+}$ was originally
measured in the inclusive $B_c^{+}\rightarrow J/\psi+l^{+}+X$ mode in CDF Run I data to be 
$6400\pm390(stat.)\pm130(syst.)$ MeV/c$^2$~\cite{Abe:1998wi}.

\begin{figure*}[t]
\centering
\includegraphics[width=100mm]{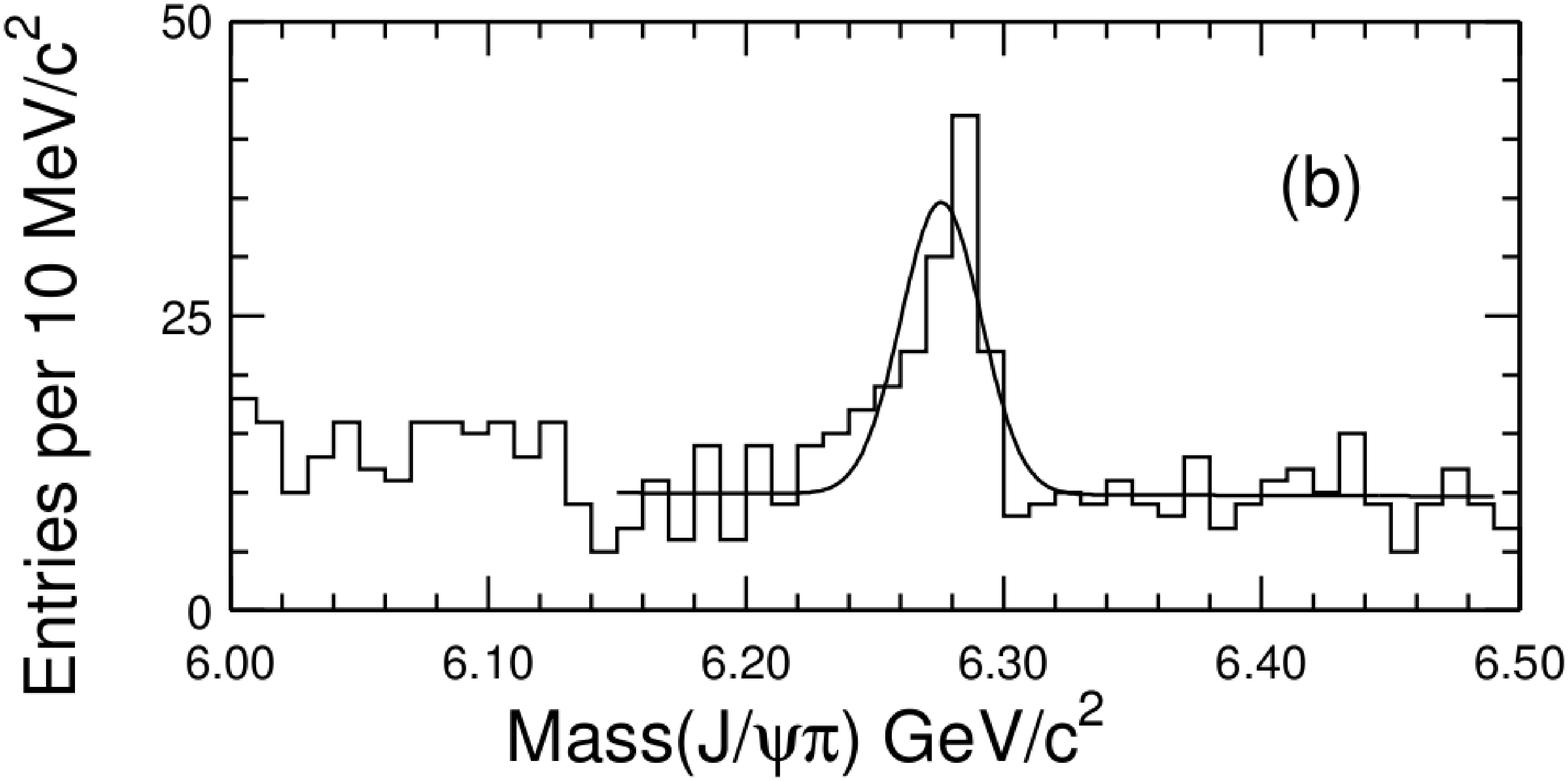} \\
\includegraphics[width=87mm]{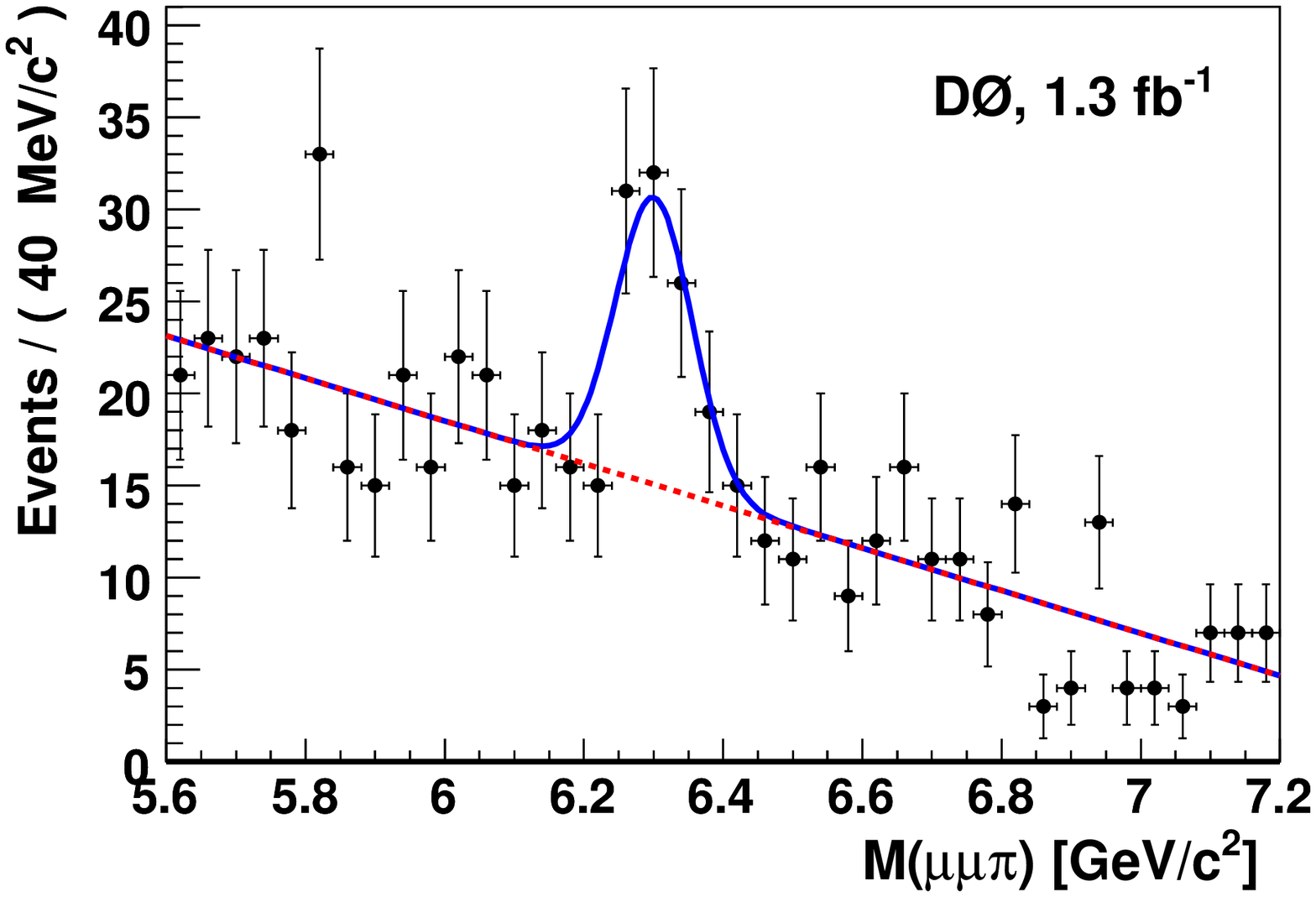}
\caption{$J/\psi\pi$ mass distributions for $B_c^{+}$ mass measurements at CDF (top) and D0 (bottom).} 
\label{bc_mass}
\end{figure*}

Both CDF and D0 measure the $B_c^{+}$ mass in the exclusive $B_{c}^{+}\rightarrow J/\psi+\pi^{+}$
decay mode, CDF with $2.4$ $fb^{-1}$ and D0 with $1.3$ $fb^{-1}$ of data.  The choice of the 
exclusive mode allows for the first measurements of the $B_c^+$ mass from fits of a fully reconstructed
mass peak.   Both analysis are based on event selection optimized for use with 
$B^{+}\rightarrow J/\psi+K^{+}$ decays.  D0 carries out an additional stage of optimization
using a sample of simulated signal events.  Fig.~\ref{bc_mass} shows the reconstructed mass distributions
for the CDF and D0 results.  CDF obtains the world best measurement~\cite{Aaltonen:2007gv}: 
\begin{eqnarray*}
M_{B_c} = 6275.6\pm2.9(stat.)\pm2.5(syst.) \: \mbox{MeV/c}^2
\end{eqnarray*}
The further improvements in precision are limited by the systematic uncertainty which is dominated by
the understanding of the detector resolution for mass measurements. D0 measures the mass as~\cite{:2008kv}
\begin{eqnarray*}
M_{B_c} = 6300\pm14(stat.)\pm5(syst.) \: \mbox{MeV/c}^2
\end{eqnarray*}
The measurement are in agreement with each other, while the CDF result may suggest refinements to 
the calculation of the mass using lQCD.

\section{LIFETIME OF THE $B_s^0$ MESON}
The $B^0_s$ meson, the ground state meson consisting of $b$ and $s$ valence quarks,
exhibits the behaviour of particle/antiparticle virtual transitions (mixing) which is seen in flavored neutral
mesons.  As a result of the mixing, the system has two mass eigenstates, $B_{L}$ and
$B_{H}$, each with its own lifetime.  Measurements of the $B^0_s$ lifetime either
differentiate the mass eigenstates and measure their lifetimes separately or measure a 
combination of the two lifetimes.  One can define the average lifetime and lifetime
difference as
\begin{equation} 
\frac{1}{\tau_{B_s}} = \Gamma_{s} = (\Gamma_{L}+\Gamma_{H})/2, \: \Delta\Gamma_{s} = \Gamma_{L}-\Gamma_{H}
\end{equation}
For flavor specific measurements, where the flavor of the $B^0_s$ is determined by the final state particles, 
the mass eigenstates are not separately measured and the measured lifetime is~\cite{Hartkorn:1999ga}
\begin{equation}
(\tau_{B_s})_{fs} = \frac{1}{\Gamma_{s}}\frac{1+(\frac{\Delta\Gamma_{s}}{2\Gamma_{s}})^2}
                                            {1-(\frac{\Delta\Gamma_{s}}{2\Gamma_{s}})^2}
\end{equation}

The theoretical estimates of lifetimes for ground state $B$ mesons containing a light 
quark can be evaluated in the heavy quark expansion (HQE).  Results are particularly precise for 
lifetimes ratios, where many theoretical uncertainties cancel.  The predicted lifetime ratio 
$\tau_{B_s}/\tau_{B_d} = 1.00\pm0.02$~\cite{Gabbiani:2004tp} shows a $2.1\sigma$ difference from the
world average measured value of $0.939\pm0.021$~\cite{hfag_bs} as of March 2007.  The lifetime 
results in the following sections will greatly improve the world average and decide whether this
discrepency is significant. 

\subsection{Lifetime in $B^0_s\rightarrow D_s^-(\pi^-\phi)+\pi^++X$ Decays}
\begin{figure*}[t]
\centering
\includegraphics[width=100mm]{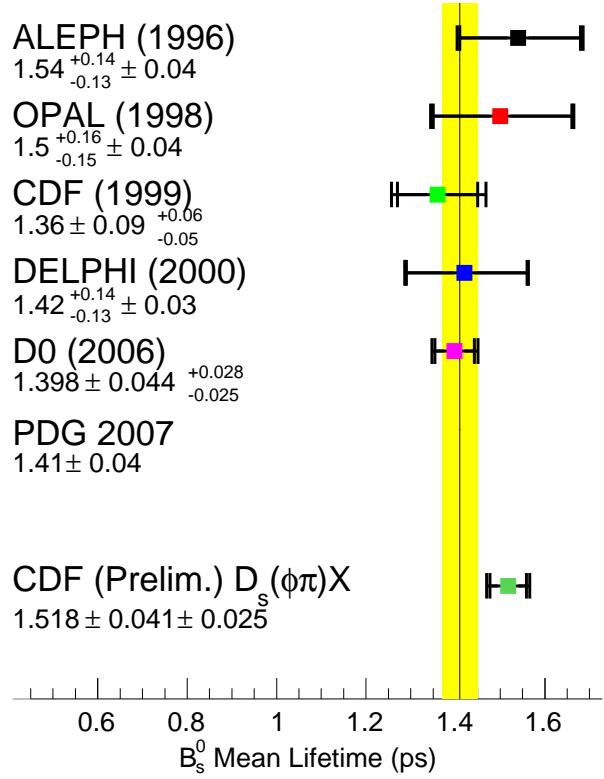}
\caption{Comparison of flavor specific measurements of the $B_s$ lifetime.} 
\label{bs_fs_sum}
\end{figure*}

CDF measures the lifetime of the $B_s^0$ in the flavor specific mode $B_s^0\rightarrow D_s^-(\pi^-\phi)+\pi^++X$,
which includes the fully reconstructed $D_s^-(\pi^-\phi)+\pi^-$ and partially reconstructed
$D_s^-(\pi^-\phi)+\rho^+$ and $D_s^{*-}+\pi^+$ final states.  The partially reconstructed modes
increase the statistics, improving the precisions, but require a K
factor from a simulation of partially reconstructed states that models the missing momentum in those
decay modes.
The mass models for the decay modes are also determined with simulated events.  The lifetime
fit takes place in two steps:  the mass distribution is fitted to constrain the fractions
of the fully and partially reconstructed decay modes, and the fractions are propagated into the 
lifetime fit.  The measured lifetime is the most precise flavor specific measurement to date~\cite{cdf_bs_fs}: 
\begin{eqnarray*}
(\tau_{B_s})_{fs} = 1.518\pm0.041(stat.)\pm0.025(syst.) \: \mbox{ps}
\end{eqnarray*}
Fig.~\ref{bs_fs_sum}, which summarises measurements of the $B_s$ lifetime in flavor specific modes,
shows that this measurement has equal precision to the previous world average and will raise the 
average considerably.

\subsection{Lifetime and $\Delta\Gamma_{s}$ in $B_s\rightarrow J/\psi+\phi$ Decays}
The average $B_{s}^0$ lifetime and $\Delta\Gamma_{s}$ is measured using the decays 
$B_s^0\rightarrow J/\psi+\phi$, where the heavy and light eigenstates can be identified.  
Since the heavy and light eigenstates are CP odd and even respectively if one neglects 
the small expected CP violation of these decays, CP of the decay
products determines the mass eigenstate.  In the decay of the pseudoscalar $B_s^0$ to the vectors 
$J/\psi$ and $\phi$, the CP of the final states is determined by the orbital angular
configuration of the $J/\psi$ and $\phi$; CP odd for P wave and CP even for S and D wave.

\begin{figure*}[t]
\centering
\includegraphics[width=84mm,height=70mm]{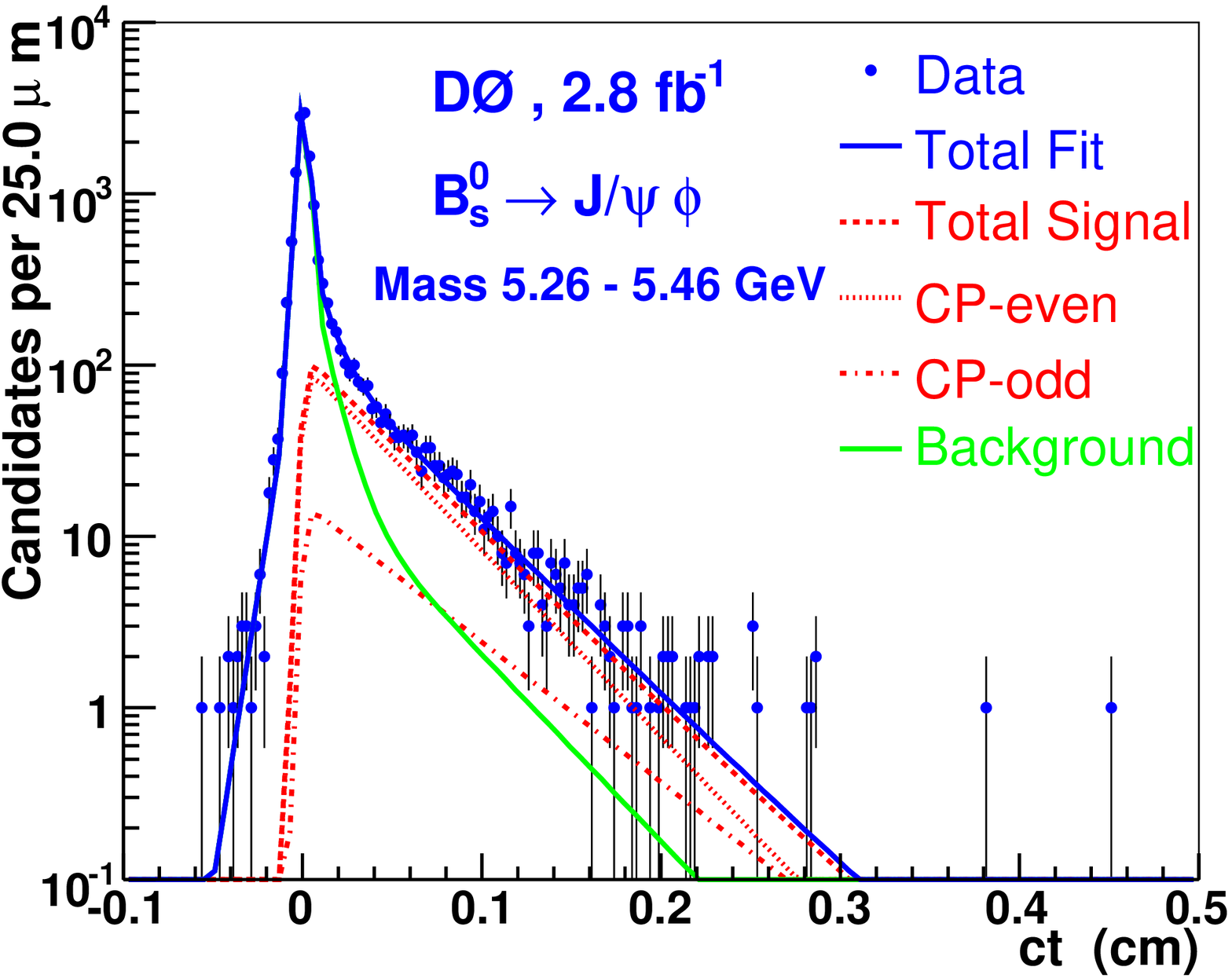}
\includegraphics[width=90mm,height=75mm]{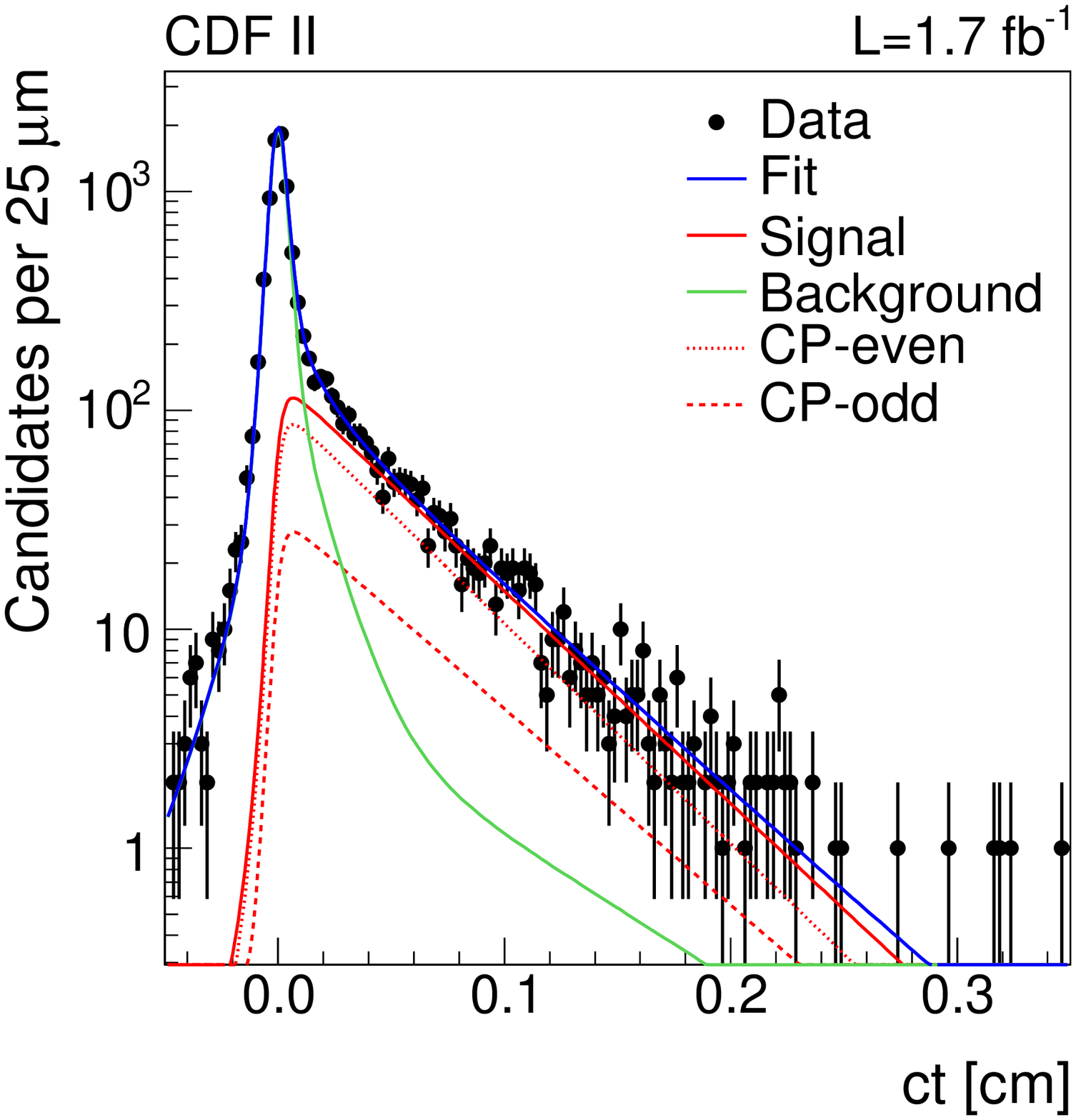}
\caption{$B_s^0$ lifetime fit projections for measurements in the $J/\psi\phi$ channel at D0 (left)
and CDF (right)} 
\label{bs_jpsiphi}
\end{figure*}

Both CDF and D0 measure the lifetime and $\Delta\Gamma_{s}$ using $1.7$ $fb^{-1}$ and $2.8$ 
$fb^{-1}$ respectively.  In the CDF measurement the CP violating phase $\beta_s$ is fixed to
0 while D0 allows it to float.  The angular component of the fit is carried out using the 
transversity basis~\cite{:2008fj} and measures the mass eigenstate contributions.  Fig.~\ref{bs_jpsiphi} shows
the lifetime projections for the D0 and CDF fits.  The fitted lifetime and $\Delta\Gamma_{s}$
from D0 are~\cite{:2008fj}
\begin{eqnarray*}
\tau_{Bs} & = & 1.52\pm0.05(stat.)\pm0.01(syst.) \: ps \\
\Delta\Gamma_{s} & = & 0.19\pm0.07(stat.)^{+0.02}_{-0.01}(syst.) \: ps^{-1}
\end{eqnarray*}
and from CDF~\cite{Acosta:2004gt}
\begin{eqnarray*}
\tau_{Bs} & = &  1.52\pm0.04(stat.)\pm0.02(syst.) \: ps \\
\Delta\Gamma_{s} & = & 0.08\pm0.06(stat.)\pm0.01(syst.) \: ps^{-1}
\end{eqnarray*}

Given the measured value of $\tau_{B_d}=1.530\pm0.009$~\cite{hfag_bs}, these results, including the 
flavor specific measurement, suggest the previous discrepancy in $\tau_{B_s}/\tau_{B_d}$ was statistical, 
and $\tau_{B_s}/\tau_{B_d}=1$ seems favored.

\section{DIRECT OBSERVATION OF THE $\Xi_{b}^{-}$ BARYON}
\begin{figure*}[t]
\centering
\includegraphics[width=80mm]{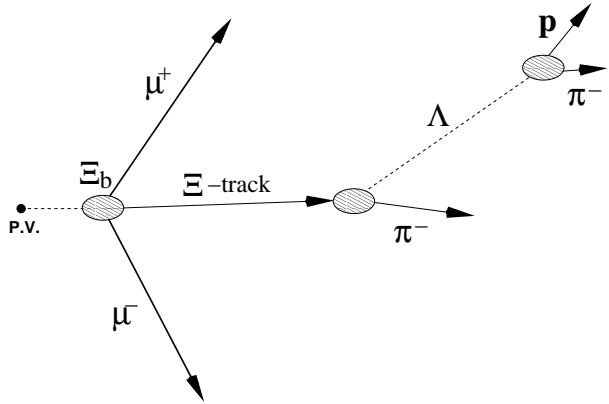}
\caption{Decay topology for $\Xi_b^-$ measurements.} 
\label{xi_top}
\end{figure*}

Until recently, direct observation of baryons containing a $b$ quark have been limited
to the $\Lambda_b$ ($udb$) baryon.  The quark model predicts a charged baryon $\Xi_{b}^{-}$
that is the $dsb$ bound state.  As an analog to $B_{s}^0\rightarrow J/\psi\phi$ decays
one will expect the $\Xi_{b}^{-}$ to decay to a $J/\psi$ and the doubly strange $\Xi^{-}$.
CDF and D0 searched for the $\Xi_{b}^{-}\rightarrow J/\psi(\mu^+\mu^-)+\Xi^{-}(\Lambda\pi^{-})$ decay chain  
(see Fig.~\ref{xi_top}) and measured the mass and relative production cross section of $\Xi_{b}^{-}$.

The selection of $\Xi_b^-$ events is driven by the relative large decay lengths of the of the $\Xi$ and
$\Lambda$ decay products, which are on the order of centimeters.  For the D0 measurement 
which uses 1.3 $fb^{-1}$ of data, large decay lengths are used by only selecting events
where the $\Lambda$ decay products have significantly large impact parameters with respect
to the $\Xi_b^-$ decay point.  In the CDF analysis, the $\Xi$ trajectory is reconstructed 
using hits in the silicon detectors, transforming a 5 track final state into a 3 track final
state and reducing the backgrounds.

D0 made the first direct observation of $\Xi_{b}^-$ with $15.2\pm4.4(stat.)^{+1.9}_{-0.4}(syst.)$
signal events at a significance of $5.5\sigma$.  D0 measures the mass and relative cross section
time branching ratio~\cite{:2007ub}:
\begin{eqnarray*}
M_{\Xi_b^-} = 5.774\pm0.011(stat.)\pm0.015(syst.) \: \mbox{GeV/c}^2 \\
\frac{\sigma(\Xi_b^-)\times\mathcal{B}(\Xi_b^-\rightarrow J/\psi\Xi^-)}{\sigma(\Lambda_b)\times\mathcal{B}(\Lambda_b\rightarrow J/\psi\Lambda)} = 0.28\pm0.09(stat.)^{+0.09}_{-0.08}(syst.)
\end{eqnarray*}
CDF observes $17.5\pm4.3$ signal events at a $7.7\sigma$ significance and measures the mass as 
well~\cite{:2007un}:
\begin{eqnarray*}
M_{\Xi^-_b} = 5.793\pm0.003(stat.)\pm0.007(syst.) \: i\mbox{GeV/c}^2 \\
\end{eqnarray*}

\begin{figure*}[t]
\centering
\includegraphics[width=100mm]{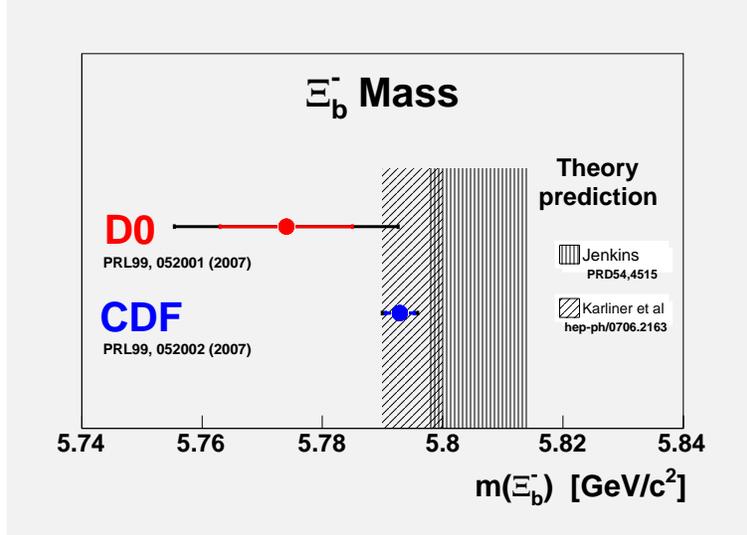}
\caption{Comparison of $\Xi_b^-$ mass measurements and theoretical predictions.} 
\label{xi_mass_sum}
\end{figure*}

Fig.~\ref{xi_mass_sum} shows the fitted mass results in comparison to theoretical 
predictions using hyperfine color interactions~\cite{Karliner:2007jp} and expansions in 
the heavy quark mass and number of colors~\cite{Jenkins:1996de}.  The measured values 
are in good agreement at the low end of predicted values.



\section{HEAVY QUARKONIUM POLARIZATION IN 1S STATES}  
The production of heavy quarkonium, $b\bar{b}$ and $c\bar{c}$ bound states,
can be understood in the framework of NRQCD~\cite{Bodwin:1994jh} where calculations
of the total cross sections are in good agreement with previous 
charmonium results from CDF~\cite{Abe:1997jz,Acosta:2004yw}.  The NRQCD approach
to heavy quarkonium production predicts that for sufficiently large
$p_T$ the $J^{PC}=1^{--}$ states should be transversely polarized~\cite{Cho:1994ih}.
Results from CDF Run I in charmonium~\cite{Affolder:2000nn} and 
bottomonium~\cite{Acosta:2001gv} do not show the predicted transverse polarization.

For measurements of the heavy quarkonium polarization, the polarization is parameterized
using $\alpha$:
\begin{equation} 
\alpha = (\sigma_{T}-2\sigma_{L})/(\sigma_{T}+2\sigma_{L})
\end{equation}
Here $\sigma_{T}$ and $\sigma_{L}$ are the transverse and longitudinal cross
sections.
For decays of the heavy quarkonium states to two muons, $\alpha$  is related to 
$\theta^*$, the angle of the positive muon in the quarkonium center of 
mass frame with respect to the quarkonium direction in the lab frame:
\begin{equation}
\frac{dN}{d(cos\theta^*)} \propto 1 + \alpha cos^2\theta^*
\end{equation} 
The value of $\alpha$ can then be determined by studying the shape of the 
$cos\theta^*$ distribution in heavy quarkonium decays.

\subsection{Polarization of the $J/\psi$}
CDF measures the polarization of $J/\psi$ and $\psi(2s)$ production as a function of $p_T>5$ 
GeV/c using $\psi\rightarrow \mu\mu$ in 800 $pb^{-1}$ of integrated luminosity.  The 
$J/\psi$ are chosen by reconstructing the dimuon mass and selecting events in a 
$3\sigma$ region around the central value. The background contribution is parameterized using events 
from $J/\psi$ mass sidebands
located $7\sigma$ from the signal region.  The contribution of $J/\psi$ from the decay of $B$
hadrons is estimated by subtracting the distribution of events with negative $ct$, where
there is only a contribution from prompt $J/\psi$ production, from those with positive $ct$,
where $J/\psi$ are produced promptly and in $B$ decays.

\begin{figure*}[t]
\centering
\includegraphics[width=100mm]{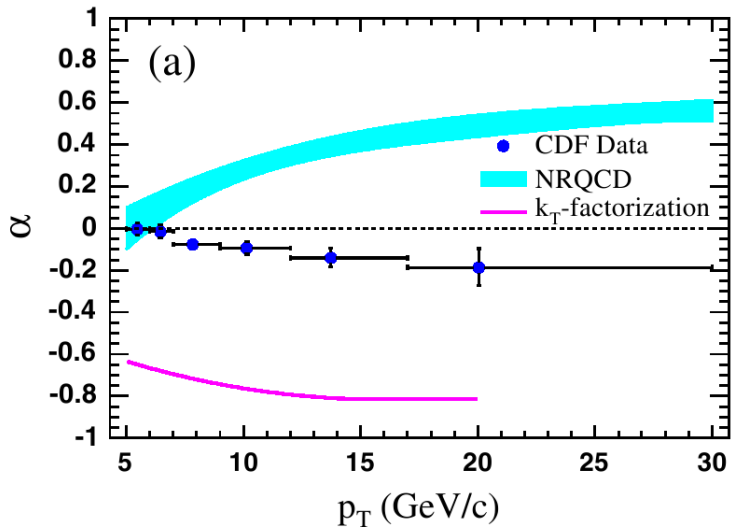}
\caption{$p_T$ dependence of the polarization parameter $\alpha$ in $J/\psi$ production at CDF.} 
\label{jpsi_polar}
\end{figure*}

The distributions of $cos\theta^*$ are fitted using templates that describe logntitudinaly and
transversely polarized $J/\psi$.   Fiq.~\ref{jpsi_polar} shows the measured polarization 
parameter $\alpha$ as a function of the $J/\psi$ $p_T$~\cite{Abulencia:2007us}.  The 
polarization parameter $\alpha$ does not exhibit the transverse dominance predicted by NRQCD.

\subsection{Polarization of the $\Upsilon(1S)$}
\begin{figure*}[t]
\centering
\includegraphics[width=100mm]{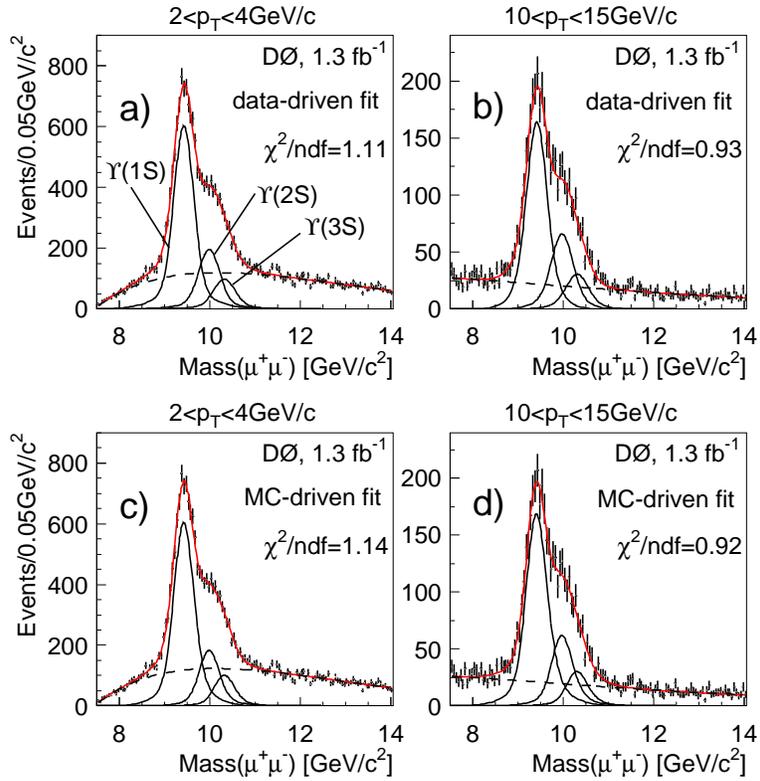}
\caption{Examples of mass fits of $\Upsilon$ states decaying to two muons.} 
\label{upsilon_mass}
\end{figure*}

\begin{figure*}[t]
\centering
\includegraphics[width=100mm]{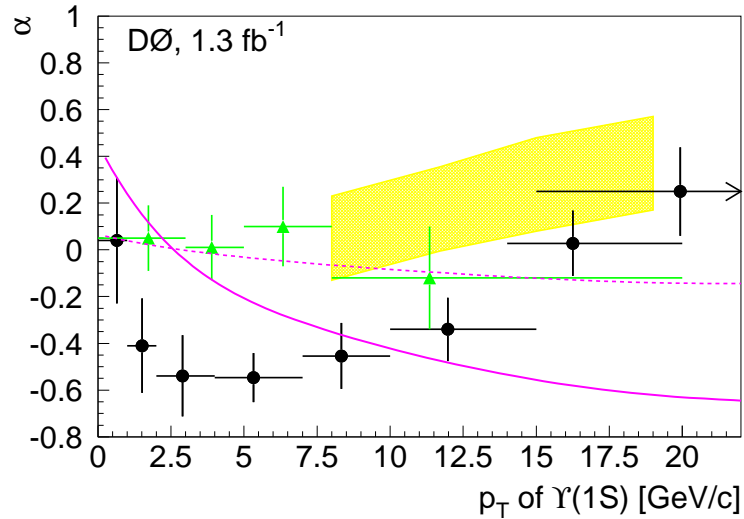}
\caption{$p_T$ dependence of the polarization parameter $\alpha$ in $\Upsilon(1S)$ production at D0.} 
\label{ups_polar}
\end{figure*}

D0 measures the polarization of $\Upsilon(1S)$ and $\Upsilon(2S)$ states in reconstructed 
dimuon pairs.  As Fig.~\ref{upsilon_mass} illustrates, the $\Upsilon$ states overlap, leading
to the importance of the mass fit in separating the contributions from the different states.
The $cos(\theta^*)$ for the transversely and longitudinally polarized states are model with
simulated events that are reweighted to match the momentum distributions for $\Upsilon$ in 
data.  The measured $p_T$ dependent value of $\alpha$~\cite{:2008za} shown in Fig.~\ref{ups_polar} does
agree with the CDF Run I measurement~\cite{Acosta:2001gv} and is in poor agreement with the 
theoretical prediction from NRQCD~\cite{Braaten:2000gw}.

\section{SUMMARY}
The CDF and D0 experiments have measured a number of properties of hadrons
containing heavy quarks that provide important feedback to the theoretical
methods used in the study of these systems.  The world best measurements of the lifetime
and mass of the $B_{c}^{+}$ as well as the heavy quarkonium polarization
test the abilities of various approaches in NRQCD to predict observables.  The
recent world best measurements of the $B^0_s$ lifetime greatly improve the precision
and show good agreement with the predictions from $HQET$.  The first direct
observation of the $\Xi_b^{-}$ suggests an exciting future in the study
of baryons containing bottom quarks.  

\begin{acknowledgments}
The author wishes to thank the organizers and hard working staff of the
Hadron Collider Physics Symposium, the Fermilab staff, and the CDF and D0 
collaborators whose efforts have yielded the results presented in 
this paper. 

Work supported by Department of Energy.
\end{acknowledgments}

\clearpage


\end{document}